\documentclass[doublecol,figures]{epl2}

\usepackage{graphicx,graphics}
\usepackage{amsmath,amssymb,epsf}

\newcommand{\de}{\delta\varepsilon}
\newcommand{\dz}{\delta Z}
\newcommand{\dphi}{\delta\phi}
\newcommand{\wst}{\omega^{*}}
\newcommand{\ws}{\omega_{s}}
\newcommand{\phic}{\phi_\mathrm{c}}
\newcommand{\Dw}{D(\omega)}

\begin{document}

\title{Excitations of Ellipsoid Packings near Jamming}

\author{Z. Zeravcic\inst{1,2}\thanks{E-mail: \email{zorana@lorentz.leidenuniv.nl}} \and N. Xu\inst{2,3} \and A. J. Liu\inst{3} \and S. R. Nagel\inst{2} \and W. van Saarloos\inst{1}}
\shortauthor{Z. Zeravcic \etal}

\institute{
 \inst{1} Instituut - Lorentz, Universiteit Leiden, Postbus 9506, 2300 RA Leiden, The Netherlands\\
 \inst{2} James Franck Institute and Department of Physics, University of Chicago, Chicago IL 60637 USA\\
 \inst{3} Department of Physics and Astronomy, University of Pennsylvania, Philadelphia PA 19104 USA
 }


\abstract{We study the vibrational modes of three-dimensional jammed packings of soft ellipsoids of revolution as a function of particle aspect ratio $\varepsilon$ and packing fraction. At the jamming transition for ellipsoids, as distinct from the idealized case using spheres where $\varepsilon = 1$, there are many unconstrained and non-trivial rotational degrees of freedom. These constitute a set of zero-frequency modes that are gradually mobilized into a new rotational band as $|\varepsilon - 1|$ increases. Quite surprisingly, as this new band is separated from zero frequency by a gap, and lies below the onset frequency for translational vibrations, $\omega^*$, the presence of these new degrees of freedom leaves unaltered the basic scenario that the translational spectrum is determined only by the average contact number. Indeed, $\omega^*$ depends solely on coordination as it does for compressed packings of spheres. We also discuss the regime of large $|\varepsilon - 1|$, where the two bands merge.}

\maketitle

\section{Introduction} There is no doubt that increasing the pressure on a floppy assembly of particles can create a rigid material. Less obvious is the assertion that such matter caused by jamming is generically different from ordinary solids. At zero temperature, frictionless, ideal spheres jam when the number of inter-particle contacts is precisely that necessary to constrain all motion; this leads to an excess of decidedly unusual low-frequency vibrational modes. However, it has been argued that frictionless spheres, with no significant rotational degrees of freedom, represent a singular situation that is too idealized to reflect the true nature of the marginally-jammed state. In particular, jammed ellipsoids can have dramatically fewer contacts than needed to inhibit all motion so that there will be many zero-frequency vibrations. Thus the distribution of collective excitations could be fundamentally altered by particle shape. Contrary to this expectation, we find the spherical jamming transition, like many singular points, controls a broader class of behavior but in an unusual non-trivial way. The new degrees of freedom introduced by ellipsoids create a rotational band but leave unaltered the basic scenario that the translational spectrum is determined solely by the average number of contacts per particle.

We study these issues in the context of the spectral density of vibrational excitations, $\Dw$. In a three-dimensional solid, the low-frequency spectrum should follow the Debye law $\Dw\sim\omega^{2}$ dictated by the elastic modes.  This is one of the most robust generic behaviors in all of materials science.  However, this law breaks down in a spectacular fashion for the case of a rigid solid formed from the jamming of spheres interacting via finite-ranged repulsions \cite{epitome,silbert05,wyartepl,wyartpre,wyartthesis,somfai,xu}. The onset of jamming in such systems has features of a first-order transition, with a discontinuity in the number of interacting neighbors per particle~\cite{durian,epitome}, as well as features of a second-order transition, with power-law scaling and diverging length scales~\cite{durian,epitome,silbert05,wyartepl,wyartpre,wyartthesis,somfai,xu,drocco,ellenbroek,olsson,hatano,bulbul,makse}.  Just above the
zero-temperature transition, $\Dw$ is approximately constant down to zero frequency \cite{epitome,silbert05} implying the existence of a new class of
low-frequency excitations that arise because the solid is on the verge of instability \cite{wyartepl,wyartpre,wyartthesis}. The Maxwell criterion for
rigidity \cite{shlomo} proposes that the average number of interacting neighbors per particle, $Z$, should be high enough to constrain all relevant
degrees of freedom in the sample: $Z \ge Z_{c}$. For frictionless spheres, the critical coordination number, $Z_{c} = 6$, {\em coincides}
\cite{epitome} with the value found at the jamming threshold packing fraction, $\phi_c$.   At packing fractions $\phi > \phi_c$, $Z$ exceeds $Z_{c}$
and consequently the plateau in the density of states persists only down to a frequency $\wst$ that depends solely on $\delta Z \equiv (Z-Z_{c})$
\cite{silbert05,wyartepl,wyartpre,wyartthesis}.  The apparent gap emerging in the spectrum between $\omega = 0$ and $\omega = \wst$ contains ordinary
elastic plane-waves described by Debye theory. Is the new physics of the excess modes robust for jamming transitions generally \cite{jamming} or is it
applicable only to this idealized situation of spheres?

It was succinctly demonstrated \cite{donev} that in one sense spheres represent a singular situation and therefore may be a poor starting point for
describing the generic properties of jammed solids. The introduction of even a small distortion to a sphere introduces many new degrees of freedom that need to be constrained for complete stability. While a sphere has only three relevant (translational) degrees of freedom, a spheroid (an ellipsoid of revolution with one symmetry axis) requires two additional coordinates (two Euler angles) to specify its orientation.  Maxwell's counting argument for the rigidity of spheroid packings would necessitate an average coordination number $Z_{c}=10$. Clearly a discontinuous increase in density would be needed if the introduction of an arbitrarily small distortion of the sphere required the average number of contacts per particle to jump discontinuously from $6$ to $10$. The rapid increase of the coordination number $Z$ with distortion and, in particular, of the packing fraction \cite{donev,donev2,man,donev3,sacanna,wouterse} has garnered much attention (``M\&M's pack more efficiently than spheres'' \cite{donev,donev2,weitz}). Nevertheless, at the jamming threshold $Z$ increases smoothly --- not discontinuously --- from $Z=6$ as spheres deform into ellipsoids, so that for small distortion $Z$ is {\em below} $Z_{c}=10$ in apparent violation of the Maxwell criterion. While there are exactly the minimum number of contacts needed for mechanical stability at the jamming transition of spheres, there are fewer than the minimum number needed for ellipsoids. There must therefore be unconstrained degrees of freedom \cite{donev3} so that the solid is not stable (to quadratic order) to some excitations. Here we investigate how rotations introduce new non-zero-frequency excitations. Remarkably we find that these modes do not destroy the picture developed for spheres but instead can be naturally
incorporated into this scenario.

\section{The Maxwell stability argument and the occurrence of zero modes}
It is clearly somewhat odd to think of $Z_{c}$ as jumping discontinuously as soon as there is a minute distortion of the particles from spherical
symmetry, so it is useful to briefly reconsider the Maxwell's argument for the stability/instability threshold \cite{shlomo}. The Maxwell criterion is
based on global counting arguments for the minimum number of contacts needed to maintain force balance on all particles. Since contacts are shared by
two particles, the minimum number of contacts necessary to clamp all particles which experience forces, $Z_{c}$, is according to this argument twice
the number of degrees of freedom of a particle. As the apparent discontinuity in $Z_{c}$ simply arises from our decision whether or not to include the
rotational degrees of freedom in the counting, it is more intuitive to restore continuity by thinking of each sphere as having 6 degrees of freedom:
three for translations and three for arbitrary rotations so that $Z_{c}=12$. The rotations of individual (frictionless) spheres do not contribute in
any way to the stability of the packing and are thus simply the $(Z_{c}-Z)/2=3$ zero-frequency modes per particle that are trivially localized on each
particle. A natural scenario is that these innocuous zero-frequency modes progressively become mobilized into finite-frequency excitations with increasing distortion of the spheres into ellipsoids. There are clearly two important values of the coordination number, $Z=6$ and $Z_{c}=12$. The important issue taken up here is the question: which of these controls the spectrum of excitations for the generalized case of ellipsoids?

From the above point of view, just above the jamming threshold, the fact that $Z$ is below $Z_{c}$ should manifest itself in the presence of
$(Z_{c}-Z)/2$ normal modes of vibration per particle with zero frequency --- see Fig.~1a. We study here the nature of these modes and the question of
how they become mobilized into finite-frequency excitations so as to find out whether this process changes the jamming scenario of frictionless spheres as the naive counting would suggest. We find that the above picture, in which the finite frequency vibrational modes are continuously turned on as the distortion is increased, unifies the scenario for aspherical particles with the one for spheres.

Since we are ignoring in the analysis below the trivial rotations about their symmetry axis, our spheroids actually have five rather than six nontrivial degrees of freedom. In this analysis, the critical contact number $Z_{c}$ is therefore $10$ rather than $12$ \cite{donev}.

\section{Preparation of packings of ellipsoids} In our numerical study, we simulate ellipsoids that are spheroidal: they have two principal axes $a$ and $b$ that are the same and a third one $c$ that is different, $a=b\neq c$. Depending on the ratio of the axes, one distinguishes between oblate ellipsoids with aspect ratio $\varepsilon = c/a < 1$ (``M\&M"s) and prolate ellipsoids with $\varepsilon = c/a >1$ (``cigars"). Particles are of equal size and mass $m \equiv 1$ and interact with the repulsive Gay-Berne potential \cite{gayberne}: $V(r_{ij},\sigma_{ij})=k/\alpha ((\sigma_{ij}-r_{ij})/\sigma_0)^{\alpha}$, where $r_{ij}=|{\bf r}_j-{\bf r}_i|$ is the distance between the centers of the ellipsoids $i$ and $j$, and $\sigma_{ij}$ is the orientation-dependent range parameter. This expression is the same as the power law potential which has been used in jamming studies of frictionless spheres \cite{epitome}: when $r_{ij}=\sigma_{ij}$ the ellipsoids just touch and for  $r_{ij}<\sigma_{ij}$ they repel with a force which is a power of the effective overlap $\sigma_{ij}-r_{ij}$; they do not interact otherwise. The range parameter $\sigma_{ij}$ is defined as
\begin{multline}
\frac{\sigma_{ij}}{\sigma_0} = \biggl[ 1 - \frac{\chi}{2}\biggl( \frac{(\hat{{\bf
r}}_{ij} \cdot \hat{{\bf u}}_i + \hat{{\bf r}}_{ij} \cdot \hat{{\bf
u}}_j)^2}{1+\chi \hat{{\bf u}}_i \cdot \hat{{\bf u}}_j} \\
+ \frac{(\hat{{\bf r}}_{ij} \cdot \hat{{\bf u}}_i - \hat{{\bf r}}_{ij}
\cdot \hat{{\bf u}}_j)^2}{1-\chi \hat{{\bf u}}_i \cdot \hat{{\bf
u}}_j} \biggr) \biggr]^{-1/2}. \label{sigmaij}
\end{multline}

Here, $\hat{{\bf u}}$ is a unit vector along the principal axis of the ellipsoid,
$\hat{{\bf u}_i}=\sin{\theta_i}\cos{\varphi_i}\,\hat{{\bf x}} + \sin{\theta_i}\sin{\varphi_i}\,\hat{{\bf y}} + \cos{\theta_i}\,\hat{{\bf z}}$, and $\chi$ is the dimensionless parameter, $\chi=(\varepsilon^{2}-1)/(\varepsilon^{2}+1)$. Spatial scales are measured in units of $\sigma_0=2a=2b$. For anisotropic particles, the parameter $k$ that appears in the expression for the potential is known as the well-depth anisotropy function and is typically taken to be a function of the three directions, $k=k(\hat{\bf u}_{i},\hat{\bf u}_{j},\hat{\bf r}_{ij})$. In order to simplify the expression for the potential, we take $k=1$ (for a one-sided harmonic potential this means that the bond stiffness equals 1 for every contact).

As in previous studies \cite{epitome}, we have used two types of interactions, harmonic, with $\alpha = 2$, and Hertzian, with $\alpha = 5/2$. We focus here on the results for the harmonic potential; results for Hertzian forces, are qualitatively the same, but will be presented in \cite{zorana09}. The configurations were made by initially placing N particles at random in a cubic box with periodic boundary conditions. We use conjugate-gradient energy minimization to obtain stable static configurations, and determine the critical jamming density $\phic$ for each of our initial configurations by incrementally compressing (decompressing) until zero pressure is reached. We then compress the system to obtain zero-temperature compressed configurations. Rattlers are removed recursively \cite{zorana09}. We study two system sizes, $N=216$ and $N=512$, and average over about 100 independent initial configurations, at each value of $Z$ and each density $\phi$. Here we will show results for $N=216$ only.

From the linearized equations of motion for the coupled translational and rotational vibrations of the ellipsoids, we construct the dynamical matrix $\mathcal D$ for each configuration. We then diagonalize $\mathcal D$, whose eigenvalues are the squared frequencies, $\omega^{2}$, of the vibrational normal modes.

\section{Analysis of the spectrum of vibrational-rotational modes}
\begin{figure}[t]
\begin{center}
\includegraphics[width=1.\linewidth]{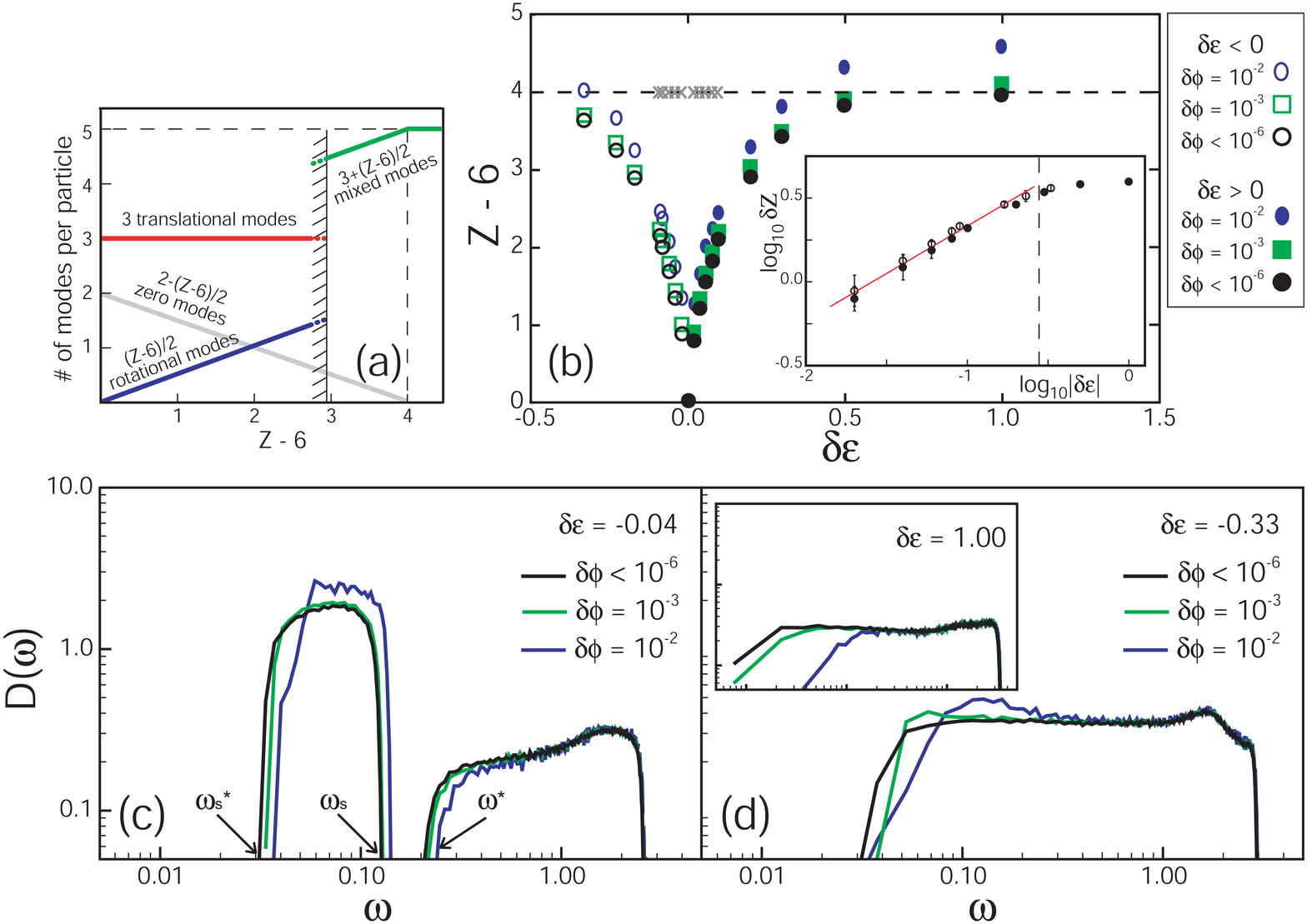}
\caption{\label{dos} The average contact number $Z$ and density of states, $\Dw$, of harmonic packings. (a) Illustration of how the number of different modes per particle (excluding rattlers) in the packings at jamming varies as a function of the average contact number $Z$. Figs.~(b-d) show that, as
the ellipticity is increased, the average contact number increases. For $Z \lesssim 9$, there are two well-defined bands: a rotational band with
$(Z-6)/2$ modes per particle and a translational band with three modes per interacting particle as is the case for spheres. Upon increasing $Z$, the
number of zero modes decreases as zero-modes are converted into finite-frequency rotational modes. Above $Z\approx 9$, there is only one band. (b) $Z$
as a function of the ellipticity $\delta \varepsilon \equiv \varepsilon -1 $ and distance from jamming $\delta \phi$ for our $216$-particle packings.
The sharp decrease around the spherical case $\delta \varepsilon=0$ is consistent with earlier results \cite{donev,sacanna,wouterse} for hard
ellipsoids and spherically capped rods. The log-log plot of $\delta Z$ versus $|\delta \varepsilon |$ in the inset shows that the rise of $Z$ at
jamming is consistent with a $\delta Z \sim \sqrt{ | \delta \varepsilon |}$ scaling. The crosses in the main plot for small values of $\delta
\varepsilon$  show that twice the measured number of zero-frequency eigenmodes per particle plus $Z-6$ add up precisely to 4, in accord with panel (a).
(c) The density of states for slightly oblate ellipsoids, $\delta \varepsilon=- 0.04$, for our packings close to jamming and at two compressions. The
existence of two bands separated by a gap as well as a gap at zero frequency is clearly visible. (d) For larger ellipticities the two bands merge, as
is illustrated here for $\delta \varepsilon= -0.33$.  The inset shows the data for $\delta \varepsilon=1$ where $Z\approx Z_{c}=10$ at jamming (see
panel a). In accord with this, the gap near zero frequency increases with increasing compression (and therefore increasing Z).  This is consistent with
the argument that $\wst$ increases as $Z $ increases above $Z_{c}$.}
\end{center}
\end{figure}

An illustration of how the number of rotational, translational and zero-frequency modes per particle at jamming varies as a function of coordination
number $Z$ is shown in Fig.~\ref{dos}a. There are $3$ translational modes per particle (red line); in the spherical limit, $Z\rightarrow 6$, there are $(10 - 6)/2 = 2$ zero energy rotational modes per particle, because the isostatic number for spheroids is 10. As $Z$ increases, the number of non-zero frequency rotational modes per particle increases as $(Z - 6)/2$ (blue line), while the number of zero modes decreases by the same amount (gray line).

In Fig.~\ref{dos}b we show the coordination number versus aspect ratio $\varepsilon$ of spheroids. The black symbols correspond to configurations evaluated very close to the jamming threshold $\phic(\varepsilon)$ for each value of the aspect ratio, $\varepsilon$. The other colors correspond to compressions $\dphi \equiv \phi - \phic$ relative to the threshold jamming density $\phic(\varepsilon)$.
Note that $Z$ depends both on $\varepsilon$ and $\dphi$. The horizontal dashed line at ($Z-6)=4$ corresponds to $Z_{c}=10$ which is the Maxwell
criterion for rigidity of spheroids. We have checked in all cases that the number of zero-frequency modes per particle at threshold is precisely
$(Z_{c}-Z)/2$ in accord with Fig.~\ref{dos}a; this is shown by the gray crosses. The inset shows that for both oblate and prolate spheroids at the threshold  $\dz \equiv (Z-6) =6.6 (3)|\de|^{0.50 (4)}$, where $\de \equiv \varepsilon -1$, in agreement with results for two-dimensional
ellipses~\cite{donev3,mailman}.

In Fig.~\ref{dos}c and \ref{dos}d, we show the averaged density of states $\Dw$ for six typical situations. In Fig.~\ref{dos}c, we show $\Dw$ for
spheroids that are close to spheres, $\de = -0.04$, for three different compressions: close to jamming at $\dphi < 10^{-6}$ (black line), at $\dphi =
10^{-3}$ (green line) and for relatively large compression, at $\dphi = 10^{-2}$ (blue line). We find that for small $\de$, the system behaves nearly
as if it was made from spheres but with a new ``rotational" band of excitations.  The plateau in the translational band of $\Dw$ still exists with a
sharp onset, $\wst$, determined by $\dz$. Our systems are too small to see the elastic plane-waves below $\wst$. As we will show, $\wst$ scales in the
same way as the plateau onset for spherical systems \cite{silbert05,wyartepl,wyartpre,wyartthesis}. The rotational band lies below the translational
band and extends over the range $\omega_s^* \le \omega \le \ws < \wst$, as marked in Fig.~1c. As we will quantify, the spectrum is therefore described as having a
lower-frequency rotational band separated by a gap from $\omega = 0$ as well as by a gap from a higher-frequency translational band.

In Fig.~\ref{dos}d, we show $\Dw$ for highly non-spherical particles, $\de = -0.33$ for the same three values of compression as shown in Fig.~\ref{dos}c. For these systems, the gap between the two bands has disappeared. Generally, the gap near $\omega = 0$ in packings which have large $\varepsilon$ and $Z\approx Z_c\ = 10$ opens up with compression above the jamming threshold. We illustrate this in the inset of Fig.~\ref{dos}d.  This is in complete agreement with the scenario for spheres \cite{epitome,silbert05}.

\begin{figure}[t]
\begin{center}
\includegraphics[width=1.\linewidth]{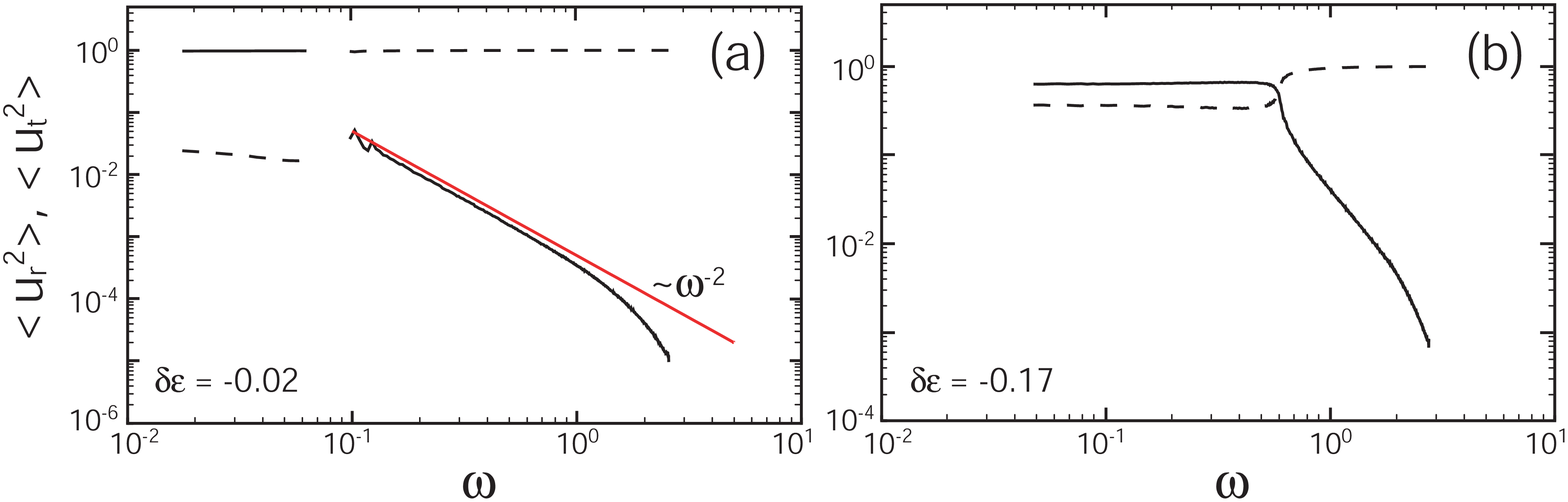}
\caption{\label{rotcontr} (a) Plot of the rotational component  $\langle u^{2}_{r}\rangle $ (solid line) and the translational component $\langle u^{2}_{t}\rangle$ (dashed line)
of the eigenmodes for $\delta \varepsilon= -0.02$ as a function of $\omega$. The lower frequency band is predominantly rotational ($\langle u^{2}_{r}\rangle \simeq 1$), while the upper frequency band is essentially translational ($\langle u^{2}_{t}\rangle \simeq 1$). The red line indicates that at high frequencies rotational contribution decreases as $\omega^{-2}$. (b) The same as in (a), but with $\delta \varepsilon=-0.17$, when the gap between the two bands has just closed and most modes have mixed character.}
\end{center}
\end{figure}

We can determine the nature of the excitations in the two bands by analyzing the eigenvectors of the dynamical matrix. We first look at the relative
contribution of the rotational degrees of freedom to the mode, $u_{\mu}(i)$, where $\mu = 1, 2, 3$ labels the translations and $\mu = 4,5$ labels the
two Euler coordinates of the orientation of each particle. In Fig.~\ref{rotcontr}, we plot the rotational contribution $\langle u^{2}_{r}\rangle =
\sum_{i=1}^{N}\sum_{\mu=4}^{5} u^{2}_{\mu}(i)/  \sum_{i=1}^{N} \sum_{\mu=1}^{5} u^{2}_{\mu}(i) $ (solid line) and the translational contribution
$\langle u^{2}_{t}\rangle = \sum_{i=1}^{N}\sum_{\mu=1}^{3} u^{2}_{\mu}(i)/  \sum_{i=1}^{N} \sum_{\mu=1}^{5} u^{2}_{\mu}(i) $ (dashed line) separately.
The lower band, existing below $\ws$, is predominantly rotational in nature while the upper band, above $\wst$, is translational. This is most
pronounced when $\varepsilon$ is small as shown in Fig.~\ref{rotcontr}a. In the limit as $\varepsilon$ approaches $0$, we find that the contribution of $\langle u^{2}_{r}\rangle$ in the upper band falls off as $(3.57(1)\cdot 10^{-4})\,\,\omega^{-2.07(1)}$ up to the onset of localized modes at high frequencies. The scaling $\sim \omega^{-2}$ is precisely what one expects from perturbation theory if the rotational degrees of freedom are weakly coupled to the translational ones. Fig.~\ref{rotcontr}b shows the data at a larger $\varepsilon$ where the bands have just merged. In this case most of the modes are of mixed character, as the rotational and translational contributions are comparable.

\begin{figure}[t]
\begin{center}
\includegraphics[width=0.5\linewidth]{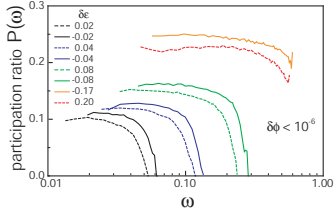}
\caption{\label{Pomega} Participation ratio $P(\omega)$ of the rotational modes for various $\varepsilon$ as a  function of frequency. The eigenmodes
at the upper edge of the rotational band are seen to be strongly localized; throughout the rest of the band $P(\omega)$ is quite flat and rather small.
The data at the largest ellipticities ($\delta \varepsilon= -0.17,$ and $0.20$) correspond to values where the gap between the two bands has just
closed --- the dip in these data is the vestige of the merging of the two bands.
}
\end{center}
\end{figure}

We can also determine the homogeneity of the modes in space by computing the participation ratio $P(\omega)= \left( \sum_{i=1}^{N}\sum_{\mu=1}^{5}
u_{\mu}^{2}(i)\right)^{2}/N\sum_{i=1}^{N}\sum_{\mu=1}^{5} u_{\mu}^{4}(i)$ of each mode. Here we will concentrate only on the lower, rotational band, since similar
studies of the translational band in spherical systems have already been reported~\cite{silbert09}. Fig.~\ref{Pomega} shows that at low values of
$\varepsilon$, the participation ratio is small and that for the highest frequencies near $\ws$ the modes become highly localized. Due to finite-size
effects, our present data must be inconclusive; however, we surmise that the rotational modes are quasi-localized for small $\de$ in the large system
limit. For the two largest values of $\de$ shown in this figure, the bands have just merged.

\begin{figure}[t]
\begin{center}
\includegraphics[width=1.\linewidth]{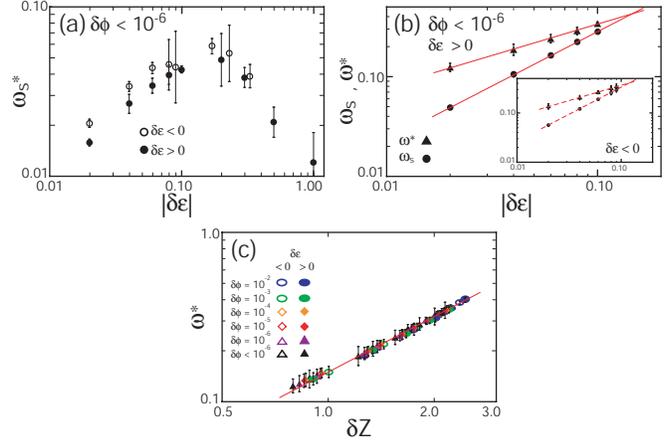}
\caption{\label{frequencies} Results for the various frequencies defined in the text and in Fig.~1c. (a) Results for $\omega^*_s$, the lower edge of the rotational band, as a function of ellipticity  at the jamming threshold. For large ellipticities $\omega^*_s$ decreases as $Z$ approaches 10. (b) Scaling of $\ws$ and $\wst$ with $\delta \varepsilon$  at the jamming threshold. Data is for prolate ellipsoids.
$\ws$, the frequency corresponding to the upper edge of the rotational band, which exists for small ellipticity, scales approximately linearly in
$\delta \varepsilon$, in agreement with the argument given in the text. Instead $\wst$ scales as $|\de|^{0.6(1)}$ (the red lines are the best
fits of the data). The point where the two lines cross marks the vanishing of the gap between the two bands. Inset: the same data for oblate
ellipsoids. Numerically, the values are very close to those for prolate ellipsoids at the same ellipticity. (c) Scaling of $\omega^*$ for various
compressions for oblate and prolate ellipsoids, showing that $\omega^*$ is determined by the contact number only, and that $\omega^*$ varies linearly
in $\delta Z$, just as it does for spheres.
}
\end{center}

\end{figure}

In Fig.~\ref{frequencies}a the frequency of the lower edge of the rotation band, $\omega^{*}_{s}$ is plotted vs. $|\de| = |\varepsilon -1|$. For small
$|\de|$, the behavior is essentially linear; for large $|\de|$, when $Z$ at jamming approaches $10$, the gap closes.

Fig.~\ref{frequencies}b shows $\ws$ and $\wst$ as functions of $|\de|$ for harmonic configurations of prolate ellipsoids that are close to the jamming
threshold. We find $\wst = 1.4 (3)|\de|^{0.6(1)}$ and $\ws = 3.5(3)|\de|^{1.1 (1)}$.
The scaling of $\ws$ can be understood as the maximum frequency of a libration mode.  As Fig.~\ref{Pomega} shows, this mode is strongly localized, so
we can obtain the scaling of the maximum frequency by estimating the torque response for rotating a single ellipsoid, keeping the other ones fixed. For a small rotation by an angle $d\theta$, a contact is compressed or decompressed by an amount $\sigma_{0}|\de|d\theta$, where $\sigma_{0}$ is the size
of the ellipsoids. This changes the normal force by an amount $k_{\rm eff}\sigma_{0}|\de|d\theta$, with $k_{\rm eff}$ the effective bond strength (for the harmonic data $k_{\rm eff} =1$). For a slightly oblate or prolate ellipsoid, the change in torque is smaller by an amount $|\de|$, hence of order $k_{\rm eff}\sigma_{0}|\de|^{2}d\theta$ as the bond vector and the normal to the surface are almost parallel. This implies that the maximum frequency $\ws\sim \sqrt{k_{\rm eff}}\, |\de|$.
Similar results are found in two dimensions~\cite{mailman}. The inset of Fig.~\ref{frequencies}b shows $\ws$ and $\wst$ as functions of $|\de|$ for oblate ellipsoids. Values of the exponents that we find for oblate ellipsoids are, within the error bars, the same as for prolate ones, with prefactors of $\ws$ and $\wst$ that are $10\%$ and $15\%$ higher, respectively.

In Fig.~\ref{frequencies}c we show $\wst$ as a function of $\dz$ for different compressions and ellipticities.  $\wst$ is still dependent solely on
$\dz$ and therefore the translational band does not depend on whether the increase in $\dz$ occurs due to an increase in compression or an increase in
the aspect ratio of the particles. For spheres, the onset of the translational band is determined by the excess number of contacts.  Our results show
that for ellipsoids, the same scenario applies, {\em irrespective of the origin of the excess contacts}. Note that the upper and lower limits of the
rotational band, $\ws$ and $\omega_s^*$ do not obey this simple behavior but depend differently on ellipticity and compression~\cite{zorana09}.

In this study we have dealt exclusively with small systems. This raises the question about what finite-size effects are important. As mentioned above, one important effect is due to plane waves. The lowest-frequency plane wave has frequency that is inversely proportional to the linear dimension of the system: $\omega_{min}= c_T k_{min}$ where $k_{min}= 2\pi/L$ with $L$ the linear system size \cite{xu}. In the systems we have studied, these are predominantly in the upper translational band\footnote{On the basis of Fig.~2 of \cite{xu}, we estimate that the frequency of an elastic mode that occurs at our largest densities $\Delta \phi$ is of order 0.3; since $\omega^*$ is almost always smaller, in our relatively small systems the translational band appears gapped.}.
In larger systems plane-wave elastic modes would begin to populate the gap between the rotational and translational band --- they would hybridize with the modes found in this study, so that localized rotational modes, for example, would become resonant or quasi-localized.

\section{Conclusions and outlook}
In conclusion, this study solves the problem of how the new degrees of freedom associated with non-spherical objects are incorporated into the
normal-mode spectrum at the jamming threshold.  Earlier findings~\cite{donev,donev2,man,donev3}  --- that the isostatic conjecture breaks down for
ellipsoid packings in the regime where $Z<10$ --- suggest that what happens for spheres does not immediately apply to more complex shapes.  As a
result, the packing problem of spheres has sometimes been viewed as an anomaly~\cite{donev,weitz}. If this extended to the nature of the jammed
state and its dynamic response, any perturbation from spherical symmetry would qualitatively change the character of marginally-jammed solid determined for spheres at Point $J$. Instead, we find that the structure of the normal-mode spectrum remains robust. The new modes that are introduced do not
affect the plateau in the density of states until the spheroid ellipticity becomes large. Moreover, these rotational modes appear to be localized so
that they should not be efficient at transporting heat. The onset of the modes in the translational band still depends only on the excess number of
contacts $\delta Z = Z-6$ as it does for spheres, irrespective of whether the excess contacts result from compression or particle asphericity.  Thus,
the singular jamming transition for spheres, Point J, in which the onset of jamming coincides with the isostatic point, controls the behavior of
systems of particles with more complex shapes, just as it controls the behavior of sphere packings that are compressed away from the transition.

There are two important regimes for ellipsoid packings: the first deals with small values of $|\de| = |\varepsilon - 1|$ where the physics is a
perturbation around the case of spheres; the second deals with large values of $|\de|$, where $Z \rightarrow Z_{c}$ (with $Z_{c}=10$ for spheroids and
$12$ for general triaxial ellipsoids). In the large $|\de|$ case, the system is well described by a theory in which a plateau in the density of states
opens up near zero frequency just as it did for the case of spheres near $Z=6$. An interesting open question is how the gap closes when $Z\rightarrow
10$ for large ellipticities.

The insights we obtain from the present study are complementary to those obtained by including friction in the vibrational spectrum of a jammed solid
\cite{somfai}.  Here we have a situation where the system jams when there are many fewer contacts than are needed according to the Maxwell rigidity
criterion, while in the case of friction there is always an excess of contacts compared to the minimum necessary for stability
\cite{zhang,makse,kasahara,somfai,shundyak}. Thus, in the case of friction there was never a question of a possible change in the underlying picture of jamming threshold. Moreover, at the Coulomb threshold to mobilization, the response dictated by the friction law is inherently discontinuous, which
makes the properties of packings with friction much more sensitive to the preparation history.

The addition of orientational degrees of freedom does introduce a new band that is essentially rotational in character while the upper band remains
nearly completely translational. It is interesting to note that the low-frequency rotational modes couple in a simple manner to the higher-frequency
translations. This is perhaps one of the root causes for the very wide spectrum of dielectric response that has been observed in glass-forming liquids
\cite{leheny}. The rotational modes should appear in the heat capacity. Indeed, the boson peak seen ubiquitously in glasses has been ascribed \cite{bosonpeak} to the excess modes associated with the plateau of $D(\omega)$ --- the fact that the jamming scenario is found to be so robust is crucial for its applicability to glasses.

\section{Acknowledgement}

We recently learned of independent parallel results by Mailman {\em et al.} \cite{mailman} with similar conclusions on two-dimensional ellipses. We
acknowledge discussions and suggestions from C. O'Hern and B. Chakraborty at the Leiden workshop ``Dynamical Heterogeneities".  In addition, we
acknowledge helpful discussions with Martin van Hecke, Silke Henkes, Brian Tighe and Tom Witten. We acknowledge the financial support of the Department of Energy: DE-FG02-05ER46199 (A.J.L., N.X.), and DE-FG02-03ER46088 (S.R.N., N.X.) and the National Science Foundation MRSEC DMR-0820054 (S.R.N., Z.Z.). Z.Z. also acknowledges support from physics foundation FOM.

\bibliography{ellipsoids}

\end{document}